\documentclass[copyright,creativecommons,noderivs,noncommercial]{eptcs}
\usepackage{amsmath}
\usepackage{amsfonts}
\usepackage{amssymb}
\usepackage{graphicx}
\usepackage{verbatim}
\providecommand{\event}{}
\providecommand{\volume}{2}
\providecommand{\anno}{2009}
\providecommand{\firstpage}{27}
\usepackage{breakurl}

\title{Towards Activity Context using Software Sensors}
\author{Kamran Taj Pathan \qquad Stephan Reiff-Marganiec
\institute{Department of Computer Science\\
University of Leicester\\ Leicester, UK}
\email{\{ktp2,srm13\}@le.ac.uk}
}
\def\titlerunning{Towards Activity Context using Software Sensors}
\def\authorrunning{K.T.~Pathan \& S.~Reiff-Marganiec}

\begin{document}
\maketitle

\begin{abstract}
Service-Oriented Computing delivers the promise of configuring and reconfiguring software systems to address user's needs in a dynamic way. Context-aware computing promises to capture the user's needs and hence the requirements they have on systems. The marriage of both can deliver ad-hoc software solutions relevant to the user in the most current fashion. However, here it is a key to gather information on the users' activity (that is what they are doing). Traditionally any context sensing was conducted with hardware sensors. However, software can also play the same role and in some situations will be more useful to sense the activity of the user. Furthermore they can make use of the fact that Service-oriented systems exchange information through standard protocols. In this paper we discuss our proposed approach to sense the activity of the user making use of software.
\end{abstract}

\section{Introduction}
The term ubiquitous was first coined by its founder Weiser~\cite{k19} in 1988 who refers it as the invisible integration of devices into everyday life. One of the fields of the ubiquitous or pervasive computing is context-aware systems. With the mobility of devices context aware systems are getting popular now a days. Pervasive computing makes use of context of the physical world, which involves a number of important concerns related to the connection of sensor information to context-aware pervasive computing, that includes: what can be feasibly sensed, the best way to acquire information and how to reason with that information to infer context~\cite{k18}.\\
It is highly needed that programs and services react specifically to their current location, time and situation and adapt their behavior according to the changing environment as context data may change dynamically~\cite{k3}\\
The information which we require can be captured through a number of ways for example by user information, network (location, time, nearby objects), sensors (activity) and other sources. One of the first context-aware applications was the Active Badge Location System~\cite{k26}, the infrared technology based system was able to determine the current location of members of the staff who wear badges and was also used to forward phone calls to a telephone near to that member. In the late 1990s some location-aware systems~\cite{k2,k7} were made and still the most frequently concerned type of context is the location. Though trying to reach the actual context many researchers have tried to find their own definition for what context actually is?\\
Schilit and Theimer~\cite{k23} used the term context-aware for the first time in 1994 (according to research papers) and described as location, identities of nearby people and objects. Brown~\cite{k5} in 1996 defined context to be the elements of the user's environment about which the computer knows. Hull et al.~\cite{k17} in 1997 described context as the aspects of the current situation.\\
So far most widely used definition is given by Dey and Abowd~\cite{k9} in 2000. They defined context as:\begin{quote}{any information that can be used to characterize the situation of an entity. An entity is a person, place, or object that is considered relevant to the interaction between a user and an application, including the user and applications themselves.}\end{quote}
Context has been classified into two categories Prekop and Burnett~\cite{k22} and Gustavsen~\cite{k13} call these external and internal, and Hofer et al.~\cite{k16} refer this as physical and logical. Physical context can be measured by hardware sensors and logical context  is mostly specified by the user or captured by monitoring the user's interactions, i.e.~the user's goals, activities, work context etc. The most relevant research area in context-aware systems make use of physical context such as location, light, sound, movement, touch, temperature etc captured by hardware sensors and for logical sensors  the Watson Project~\cite{k6} and the IntelliZap Project~\cite{k10} which provide relevant information due to information read out of opened web pages, documents etc~\cite{k3}\\
It has been observed that much of hardware and software is already installed at the organizational level and billions of machines are connected to each other; despite of that much concern is given to hardware sensors~\cite{k7} as far as the information about user's context is concerned. Hardware sensors may be more expensive and take time to install. Software can be used in addition to the hardware or on their own to sense the context, when utilized properly. This is made even more productive if we consider communicating software as well as the put over of functionality now available as services.\\
In the light of above paragraphs, we can easily infer that the most researched type of context is location and less work is done in activity context. While sensing activity currently the preferred way is to use physical sensors (e.g.~standing, walking, sitting, typing etc.) and less is logical/virtual sensors e.g.~emailing, supervising, administering etc.) which can go some steps beyond to sense the activity and can respond a user as per her own needs and criteria, which is the cornerstone to the field of context-aware systems.\\
Apart from hardware sensors (physically) we can obtain the information virtually by user's own information and can infer the outcome by applying reasoning rules.\\
In this paper Section 2 discusses related work. In Section 3 we explain the proposed context model for context information. In Section 4 overview of the architecture in association with conceptually layered framework of existing systems and our proposed approach is given. And finally Section 5 concludes and provides some future work.
\section{Related Work}
A well designed model is a key to exploiting context in any context-aware system. In this section we portray existing context models used for representing, storing and exchanging context information. Strang and Linhoff-Popien~\cite{k24} provide a survey of models with respect to software sensors in which they also introduce a classification of models based on the used data structures.\\
Schilit et al.~\cite{k23} used key-value pairs to model the context by providing the value of a context information on location to an application as an environment variable. Key-value pairs are easy to manage, but lack capabilities for sophisticated structuring for enabling efficient context retrieval algorithms.\\
Markup Scheme models are based on a hierarchical data structure which consists of markup tags with attributes and content. Specifically, the content of the markup tags is defined by other markup tags. Representatives of this kind of context modeling approach are profiles. They are usually based upon XML type languages such as RDF/S and have the advantage of easy tool access, but lack of formality and expressiveness.\\
The Unified Modeling Language (UML) has a graphical component (UML diagrams). Due to its  standard structure, UML has been used to model context. Bauer~\cite{k4} and Henricksen et al.~\cite{k14,k15} modeled contextual aspects relevant to air traffic management using UML extensions.\\
In a logic based context model, the context is described as facts, expressions and rules. Usually contextual information is added to, updated in and deleted from a logic based system in terms of facts or inferred from the rules applied on the system. A first logic based context modeling approach has been published by McCarthy et al.~\cite{k20}. McCarthy and others introduced contexts as abstract mathematical entities with properties useful in artificial intelligence.\\
An ontology is a mechanism to specify concepts and their interrelations~\cite{k11}. Context models based on ontologies have been first proposed by Otzturk and Aamodt~\cite{k21}. They derived the necessity of normalizing and combining the knowledge from different domains. Ontologies enable contextual knowledge sharing and reuse in a ubiquitous computing system. This contextual knowledge is evaluated using an ontology reasoner. Another context modeling approach based on ontologies is the CoBrA system~\cite{k8}. CoBra provides a set of ontological concepts to characterize entities such as persons, places or several other kinds of objects within their contexts and uses a broker-centric agent architecture.\\
While early models mainly addressed the modeling of context with respect to one application or an application class, generic context models are of interest since many applications can take benefit from these and can share knowledge across different systems. Though the model-oriented approach supports formality, context reasoning is usually based on Semantic Web technologies. Gu et al.~\cite{k12} have modelled context based on an ontology-oriented approach but this lacks the upper ontology by not making it more general, which may affect context reasoning in the domain specific ontologies. In this paper we have extended their work and present our ontology-based context model that addresses these shortcomings. We have emphasized more on activity context because to describe a rich source of information for advanced adaptation in collaboration like inContext~\cite{k25}, activity context plays a major role.\\
We have chosen Web Ontology Language (OWL) because it is very expressive compared to other ontology languages, it has the capability to be distributed across different systems, scalability to web needs, compatibility with web standards for accessibility and internationalization, openness and extensibility, and enabling automated reasoning to be used by automated processes and of all a W3C standard~\cite{k1}. Also being a Web language it is an obvious choice to be used in connection with services.
\section{An Ontology-based Model}
In this section we describe our design considerations together with an activity-aware meeting scenario to be used to demonstrate our context model.
\subsection{An Activity-aware Meeting Scenario}
\textbf{John:}~A Faculty~~~~\textbf{Jim and Kim:}~Research Students\\

\noindent An activity-aware meeting makes use of software sensors to sense the activity of the user. In this section, we describe a typical scenario in order to demonstrate our modeling concept.\\

\noindent Jim, a research student, wants to meet with his supervisor, John. When he looks at John's current activity, it is showing that John is Teaching because John's timetable service has updated his current activity.\\

\noindent John, a Researcher is sending an email to one of his research students, Kim, to fix a meeting on Tuesday 11:00 at his office, and the Email service has updated his context on that particular time as 'Meeting'.\\

\noindent John wants to book a trip to a conference and before booking, updated his current context as 'Out for Conference', but when he goes through the weather service for the weather condition on that day, he realizes that trip is not possible due to weather condition, and the context is updated back as planning for trip.
\subsection{Design Considerations}
The context model should support semantic interoperability to enable the common schemas to be shared between different entities because a context-aware system requires information to be shared and used between different entities such as users and services. In the above scenario the representation of John's activity should be understood from his personal profile, calendar, timetable and email services.\\
According to the definition given by Dey and Abowd~\cite{k9} context has a great variety. Context information varies in different domains. We are more concerned about activity context here, which can also affect the other factors of context. Context information is interrelated as per the above definition for example in our scenario if John has updated his calendar for a trip but because of weather condition he is unable to fly then we will have to consider this factor also.\\
In this discussion we have come to know that ontology based modeling is the approach which uses proper knowledge management, avoids inconsistency, and applies reasoning rules. The beauty of this approach is that in the future context sources become reusable and extendable. For software sensors to work without any conflict these features play a vital role.
\subsection{Context Ontology}
The basic concept of our context model is based on ontology inspired by Gu et al.~\cite{k12}, which provides a vocabulary for representing knowledge about a domain and for describing specific situations in a domain. We have chosen context ontology because it defines a common vocabulary to share context information, share common understanding of the structure of context information among users devices and services to enable semantic operations, includes machine-processable definitions of basic concepts in the domain and relations among them and reasoning becomes possible by explicitly definition. Since we use software as sensors these characteristics help to achieve our target.\\
The ability of the context ontology is to capture all the characteristics of context information which is a very difficult task in a context-aware environment. As the domain of context can be divided  into  types and further into sub-domains, it would be easy to specify the  context  in  one  domain  in  which  a  specific  range  of context  is  of  interest. The  separate processing of a proper domain  may also reduce  the  burden  of  context  processing  and  make  it possible  to  interpret  context  information  on  a variety of devices for example PC and PDA etc.\\
Our context ontologies are divided into high-level ontology and low level ontologies. The  upper ontology  captures  general context  knowledge  about  the  physical  world  in  context-aware computing  environments.  The  low-level ontologies define  the details of general concepts and  their properties  in each sub-domain. With the change of an environment the domain-specific ontology  can be dynamically  plugged  into  and  unplugged  from  the  upper ontology. For  example, when a user is sending email then it switches to an email service.
\begin{figure}[h]
   \begin{center}
   \includegraphics[width=\textwidth]{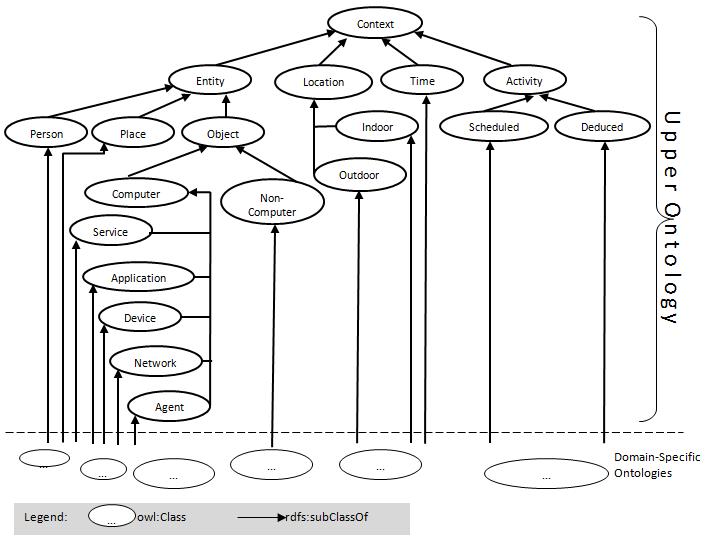}
   \end{center}
  \caption{Class hierarchy diagram for context ontologies inspired from Gu et al.~\cite{k12}.}
  \label{fig1}
\end{figure}
To make the upper ontology more generic, we have classified the context into four main types, the Entity, Location, Time and Activity as shown in Figure 1. The class Context provides an entry point of reference for declaring the upper ontology, one instance of context exists for each user or service. Each instance of Context presents a set of descendant classes of Entity, Location, Time and Activity.\\
The details of these basic concepts are defined in the low level ontologies, which vary from one domain to another. We will define all the descendent classes of these basic classes in activity-aware meeting environment and a set of properties and relationships that are associated with these classes.
\subsection{Context Reasoning}
One of the main features of the context model is the ability to support the process of reasoning about various types of context and its properties. We consider two types of Activity context. First is scheduled activity and the other is deduced activity. Reasoning in Context broadens context information implicitly by introducing deduced context derived from other types of context and also provides a solution to resolve context inconsistency that is caused by unsatisfactory sensing~\cite{k12}.\\
By reasoning context, scheduled context can be inferred from defined context but deduced context can be inferred from sensed, planned and or aggregated context based on our classification. For example in our scenario, scheduled context John's current activity can be inferred from sensed context John's calendar and timetable but deduced context can be inferred from sensed context John's calendar, timetable, email, weather as illustrated below:\\

Timetable(John, Office) $\wedge$  Calendar (John, Personal) $\vdash$ Teaching (John, Class)\\

\noindent The above example shows that, If John's office timetable and personal calendar states the particular time is scheduled to conduct a class infers John's activity as 'Teaching'\\
A more complicated example below shows deduced context where John has updated his context as 'Out for Conference' but because of poor weather conditions (snow in this case) his context has been changed to 'Planning for Conference'.\\

Search(John, Flight) $\wedge$  WeatherCond(Weather, Snowing) $\vdash$ Activity (John, NO)\\

\noindent By reasoning context classification information based on our context model, we may be able to detect and resolve context conflict. For example if John is switching between activities like 'Meeting', 'Discussing on Project' and 'Presenting' then we may resolve that John is 'Meeting for Project'. Different types of context have different levels of confidence and reliability. For example scheduled activity is more reliable than deduced activity thus varies in percentages.
\section{Architecture Overview}
In this section, we describe the basic service-oriented middleware architecture for the proposed system with an association of conceptually layered framework. Context-aware systems can be implemented in many ways. The approach depends on special requirements and conditions such as sensors (software or hardware), the user (one or many), the available resources of the used devices (PCs or mobile) or the facility of a further extension of the system. Furthermore, the method of context-data acquisition is very important when designing context-aware systems because it predefines the architectural style of the system at least to some extent~\cite{k3}. This architecture consists of the following components as shown in Figure 2:

\begin{figure}[h]
  \begin{center}
  \includegraphics[width=.95\textwidth]{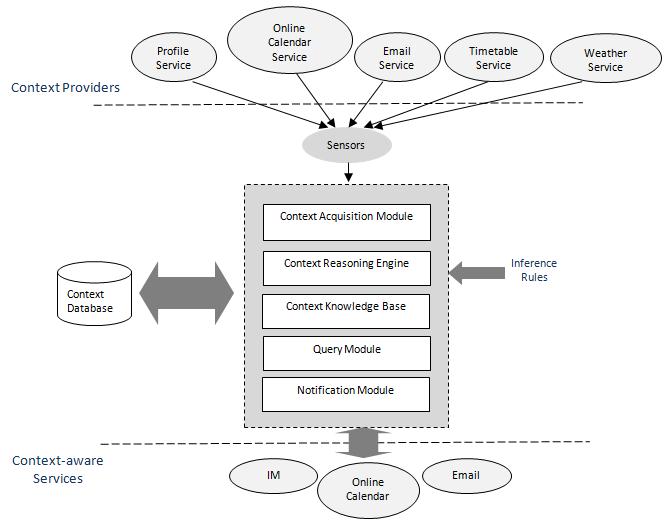}
  \end{center}
    \caption{This system acquires context information of the user from sensors in its environment and infer the current context.}
    \label{fig2}
\end{figure}

\begin{itemize}
  \item \textbf{Context Providers:} Context Providers are the services which provide the user information such as profile, calendar, timetable etc.
  \item \textbf{Sensors:} Sensors here means software sensors which are used to extract context information from different services to context acquisition module.
  \item \textbf{Context Acquisition Module:} The Context Acquisition Module acquires context from the different services and present them to OWL representations so that context can be shared and reused by other components.
  \item \textbf{Context Reasoning Engine:} The Context Reasoning Engine provides the context reasoning services including inferring deduced contexts, resolving context conflicts and maintaining the consistency of the context knowledge base.
  \item \textbf{Context Knowledge Base:} The Context Knowledge Base provides the service that other components can query, notify, add, delete or modify context knowledge with the help of Query Module and Notification Module stored in the context database.
  \item \textbf{Context-aware Services:} Context-aware services make use of different levels of context and adapt the way they behave according to the current context.
\end{itemize}

\noindent Based on this architecture, we will implement a prototype that aims to realize the activity-aware meeting scenario that is described in section 3.
\section{Conclusion and Future Work}
In this paper, we have described an approach to sense the activity of user with the help of software sensors by acquiring and applying reasoning rules to infer the context. The software sensors are used to manage exchanges occurring on service invocation. Frequently, used and tested sources for activity sensing are hardware sensors and we have tried to alter this to software sensors, which are also capable of sensing activity context. In this approach we have also emphasized that by using semantic web technologies proper knowledge management can be achieved and thus context sources become reusable and extendable. We have also proposed a more generic and extensible context model to provide a vocabulary for representing knowledge about a domain and for describing specific situation.\\
To demonstrate the feasibility of the proposed research the future work in this area will be to build a prototype of activity-aware context monitoring, to infer the activity of a user.
\section*{Acknowledgments}
The authors thank Dr. Imdad Ali Ismaili, Dr. Monika Solanki and Yi Hong for their valuable discussion and also acknowledge University of Sindh, Pakistan for scholarship No.SU/PLAN/F.SCH/323/2007.
\bibliographystyle{eptcs}
\bibliography{tacuss}
\end{document}